\PassOptionsToPackage{dvipsnames,table,xcdraw}{xcolor}
\documentclass{article}

\PassOptionsToPackage{numbers, numbers,sort&compress}{natbib}

\usepackage[preprint]{neurips_2024}

\usepackage[utf8]{inputenc} % allow utf-8 input
\usepackage[T1]{fontenc}    % use 8-bit T1 fonts
\usepackage{hyperref}       % hyperlinks
\usepackage{url}            % simple URL typesetting
\usepackage{booktabs}       % professional-quality tables
\usepackage{amsfonts}       % blackboard math symbols
\usepackage{nicefrac}       % compact symbols for 1/2, etc.
\usepackage{microtype}      % microtypography

\usepackage{enumitem}
\usepackage{amsmath, amssymb}
\usepackage{algorithm}
\usepackage{algpseudocode}
\usepackage{booktabs}
\usepackage{multirow}
\usepackage{graphicx}
\usepackage{url}
\usepackage{csquotes}
\usepackage{tikz}

\usepackage{xcolor}
\usepackage{caption}
\usepackage{subcaption}
\usepackage{pgfplots}

\usepackage{pgfplotstable}
\pgfplotsset{compat=newest}
\usepackage{wrapfig}

\definecolor{shallowblue}{RGB}{200, 220, 255}

\title{Inference-time Scaling for Diffusion-based \\ Audio Super-resolution}

\author{%
  Yizhu Jin\textsuperscript{1}, Zhen Ye\textsuperscript{1}, Zeyue Tian\textsuperscript{1}, Haohe Liu\textsuperscript{2}, Qiuqiang Kong\textsuperscript{3}, Yike Guo\textsuperscript{1}\footnotemark[1]\thanks{Corresponding author.}, Wei Xue\textsuperscript{1}\footnotemark[1]\\
  \textsuperscript{1}The Hong Kong University of Science and Technology \\
  \textsuperscript{2}University of Surrey\\
  \textsuperscript{3}Chinese University of Hong Kong
}

\begin{document}

\maketitle

\begin{abstract}
Diffusion models have demonstrated remarkable success in generative tasks, including audio super-resolution (SR). In many applications like movie post-production and album mastering, substantial computational budgets are available for achieving superior audio quality. However, while existing diffusion approaches typically increase sampling steps to improve quality, the performance remains fundamentally limited by the stochastic nature of the sampling process, leading to high-variance and quality-limited outputs. Here, rather than simply increasing the number of sampling steps, we propose a different paradigm through inference-time scaling for SR, which explores multiple solution trajectories during the sampling process. Different task-specific verifiers are developed, and two search algorithms, including the random search and zero-order search for SR, are introduced. By actively guiding the exploration of the high-dimensional solution space through verifier-algorithm combinations, we enable more robust and higher-quality outputs. Through extensive validation across diverse audio domains (speech, music, sound effects) and frequency ranges, we demonstrate consistent performance gains, achieving improvements of up to 9.70\% in aesthetics, 5.88\% in speaker similarity, 15.20\% in word error rate, and 46.98\% in spectral distance for speech SR from 4\,kHz to 24\,kHz, showcasing the effectiveness of our approach. Audio samples are available at: \url{https://racerk.github.io/tt-scale-audiosr/}.
\end{abstract}

\section{Introduction}

Audio super-resolution (SR) aims to estimate high-frequency components from a low-resolution (LR) audio signal, thereby expanding its bandwidth and enhancing perceptual quality. Generally, audio SR is an ill-posed task, as the missing high-frequency content cannot be uniquely inferred from the observed low-frequency signal. In practice, this manifests as a one-to-many mapping problem: a single LR input may correspond to multiple plausible high-resolution (HR) outputs~\cite{liu2021voicefixer, ye2023comospeech, yu2023conditioning, lee2024noise}. Deterministic models, which produce a single output that typically regresses to the average of all plausible HR outputs~\cite{wang2018esrgan, lee2024noise}, 
often fail to capture the mapping dynamics, limiting their ability to generate consistently high-quality HR predictions.

Recent advancements in diffusion models have shown great promise in modeling the distribution of high-dimensional data such as audio waveforms and spectrograms \cite{chen2024musicldm, liu2023audioldm, shen2023naturalspeech, ye2024flashspeech, ye2023comospeech}. Unlike traditional discriminative models, diffusion models learn to reverse a forward noise process, sampling from a Gaussian noise and iteratively transforming it into realistic data sample. Early works like NU-Wave \cite{lee2021nu} pioneers the use of diffusion model for audio SR, while NU-Wave2 \cite{han2022nu} improves upon this by incorporating short-time Fourier convolution for more effective spectral modeling. 
To better handle mismatches between training and testing bandwidths—common in real-world scenarios where audio may be captured from diverse environments or devices with varying frequency responses, compression artifacts, or sampling rates, several models including NVSR~\cite{liu2022neural}, NU-Wave2~\cite{han2022nu}, and AudioSR~\cite{liu2024audiosr} support flexible input bandwidths. Among them, AudioSR is notable for generalizing beyond the speech domain to more diverse audio content such as music and sound effects.
 
However, existing approaches have largely overlooked the role of inference-time randomness and uncertainty~\cite{han2022nu, lee2021nu, liu2024audiosr}. Diffusion models inherently introduce sampling stochasticity, generating different HR outputs from the same LR input. Although this randomness is fundamental to the generative process~\cite{ho2020denoising, song2020score, lu2022dpm}, its impact on perceptual audio quality has remained largely uncontrolled.
In practice, we observe that perceptually similar LR signals can correspond to semantically distinct HR waveforms. For instance, in speech, different timbres or phonemes may be downsampled into nearly identical LR representations. Reversely, naive SR models often generate HR outputs with mismatched generation targets, which degrade intelligibility or alter speaker characteristics—evidenced by increased \emph{Word Error Rate} and lower \emph{Speaker Similarity} scores. Similarly, in music and sound effects, low-frequency components alone may not convey the full semantic content of the original signal, resulting in HR predictions that diverge from the intended meaning.
These degradations in task-specific attributes call for targeted strategies during diffusion sampling process to restore lost attributes without retraining the model to retain both diversity and accuracy.

Recently, the study of Large Language Models has shown that allocating more computational resources at inference time through sophisticated search strategies can yield higher-quality and more contextually appropriate responses. This concept, known as \emph{inference-time scaling}, highlights a promising direction for enhancing model performance without altering the training process \cite{brown2024large, snell2024scaling, ye2025llasa}. Analogously, inference-time scaling has been explored in diffusion models for vision tasks, where increasing the compute budget beyond denoising steps, has been shown to improve generative quality ~\cite{ma2025inference, xie2025sana, zhang2025inference, zhang2025context}. 

Inference-time scaling techniques remain largely underexplored in the audio domain for diffusion models. In this work, we introduce a unified framework that applies inference-time scaling to improve audio SR quality by increasing the compute budget during inference. Specifically, our method samples multiple HR candidates using inference-time search algorithms and evaluates them using task-specific search verifiers, and selects the best-performing outputs. This approach allows us to navigate the solution space more effectively and recover critical perceptual qualities lost during vanilla SR generation.
During this process, we find that over-optimizing with a single verifier can lead to overfitting and unintended artifacts \cite{clark2023directly, pan2022effects}. To mitigate this, we ensemble multiple verifiers with complementary goals, enabling better trade-offs and more reliable improvements across diverse metrics.

Beyond performance gains, we quantify the range of the search space induced by different algorithms and examine the sample-wise variability of the diffusion process via uncertainty estimation. % These analyses provide insights into the stochastic nature of inference and its role in audio generation quality.

Our key contributions are summarized as follows:
\begin{itemize}[leftmargin=*]
    \item We present the first systematic study of inference-time scaling for diffusion models in the audio domain, introducing a general framework that combines verifier-guided search with scalable compute algorithms to enhance perceptual quality across diverse audio types for audio SR task.
    \item We analyze verifier hacking effect and employ a verifier ensembling strategy to mitigate it, enabling better trade-offs across evaluation metrics. Our analysis further reveals that different task and upsampling settings exhibit distinct preferences for specific verifier–algorithm configurations.
    \item We quantitatively characterize the range of search space induced by different search algorithms, and perform uncertainty estimation of individual samples to reveal stochastic dynamics in the diffusion process for audio SR.
\end{itemize}

\section{Related Work}
\label{sec:bg}

Denoising Diffusion Probabilistic Models (DDPMs) have emerged as a leading class of generative models capable of producing high-fidelity outputs across a variety of domains, including images, 3D, audio, and video~\cite{ho2020denoising, song2020denoising, rombach2022high, guo2023animatediff, liu2023audioldm, shen2023naturalspeech, tian2025audiox, wu2025difix3d+, zhang2025flashvideo}. DDPMs define a forward diffusion process that gradually corrupts clean data by adding Gaussian noise over $T$ steps, and learn to reverse this process via a parameterized denoising network.

The original DDPM sampling process is computationally expensive due to its many iterative steps. To address this problem, Denoising Diffusion Implicit Model (DDIM)~\cite{song2020denoising} introduces a deterministic, non-Markovian alternative to reverse sampling, enabling much faster generation while maintaining quality. Furthermore, to enable conditional generation without the need for explicit supervision, classifier-free guidance~\cite{ho2022classifier} allows the model to trade off between fidelity and conditioning strength by interpolating between unconditional and conditional predictions during inference.

As diffusion models scale to more complex data, such as high-resolution audio waveforms, efficiency becomes even more critical. Latent Diffusion Models (LDMs)~\cite{rombach2022high} address this challenge by moving the generative process into a learned latent space. A variational autoencoder first compresses the data into a lower-dimensional representation where diffusion is more efficient, and a decoder then reconstructs the signal back into its original form. In the audio domain, this approach proves especially effective due to the high temporal resolution and redundancy of waveform data. Instead of modeling raw waveforms directly, recent methods generate high-resolution mel-spectrograms in latent space, which are then converted into waveforms using neural vocoders~\cite{ren2020fastspeech, ren2019fastspeech, ye2023comospeech}. This two-stage architecture achieves both computational efficiency and high perceptual quality.

\section{Methodology}
\label{sec:method}

\subsection{Overview}

\begin{figure*}[tb]
  \centering
  \includegraphics[width=0.9\linewidth]{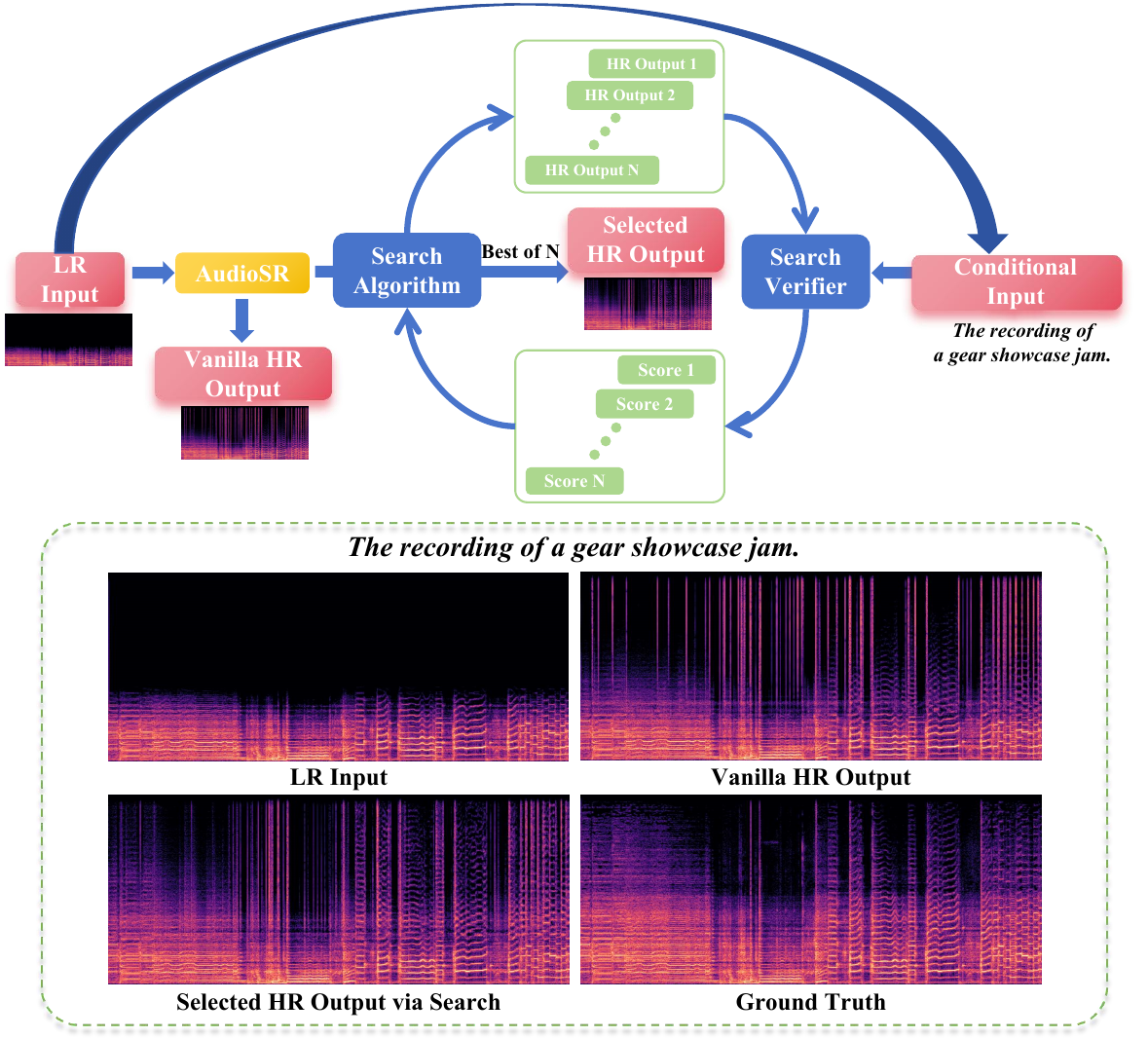}

  \caption{Overview of our inference-time scaling framework for audio SR. Given a LR input, multiple HR candidates are generated via diffusion sampling. A search algorithm explores this candidate space, guided by a verifier that scores each output based on a reference or task-specific criterion. The bottom row presents detailed STFT spectrograms of a music example. The selected output better aligns with the conditional text description and exhibits structural patterns closer to the reference.}
  \vspace{-3mm}
\label{fig:intro}
\end{figure*}

We propose a unified inference-time scaling framework for audio SR with diffusion models, as illustrated in Figure~\ref{fig:intro}. Our approach systematically explores the generative search space at inference by generating multiple HR audio candidates from an LR input and selecting the most promising output according to task-specific criteria. This is achieved by integrating two key components: (1) \emph{search verifiers}, which evaluate the perceptual or semantic quality of each candidate, and (2) \emph{search algorithms}, which efficiently traverse the candidate space guided by verifier feedback. 

For the underlying audio SR model, we adopt the state-of-the-art AudioSR~\cite{liu2024audiosr}, which is based on LDM tailored for audio. AudioSR first predicts the HR Mel spectrogram conditioned on the LR input, and then reconstructs the waveform using a pretrained HiFiGAN-based vocoder~\cite{kong2020hifi, you2021gan}. AudioSR supports a wide range of cutoff frequencies and audio types. 

Following \cite{ma2025inference}, our inference-time scaling framework is structured along two principal axes:

\textbf{Verifiers} are pretrained evaluation modules that assign scalar scores to generated HR audio, reflecting their quality with respect to specific tasks or conditions. Formally, a verifier is defined as
\begin{equation}
\mathcal{V}: \ \boldsymbol{X}^{T} \times \boldsymbol{C}^d \rightarrow \mathbb{R},
\end{equation}
where \( \boldsymbol{X}^{T} \) denotes the HR waveform of length \(T\), and \( \boldsymbol{C}^d \) is an optional conditioning input (e.g., text, transcript, or audio prompt). Verifiers guide the ranking and selection of candidates, steering generation toward outputs that better satisfy task objectives.

\textbf{Search Algorithms} leverage verifier scores to identify the best HR candidate from a set of generated samples. Formally, a search algorithm is represented as
\begin{equation}
    f: \ \mathcal{V} \times \mathcal{D}_{\theta} \times \{ \boldsymbol{X}^{T} \}^N \times \boldsymbol{C}^d  \rightarrow \boldsymbol{X}^{T},
\end{equation}
where \( \mathcal{D}_{\theta} \) is the pretrained diffusion-based AudioSR model, and \(N\) is the number of candidate HR waveforms generated per inference. The algorithm selects the candidate with the highest verifier score, enabling systematic comparison across different verifier–algorithm configurations. For fair evaluation and efficient searching, we fix $N$  during inference-time search for each method.

We further detail the search verifiers and search algorithms designed for audio SR as below.

\subsection{Search Verifiers} 

We categorize search verifiers into two classes: \emph{Oracle Verifier} and \emph{Supervised Verifier}, based on their access to privileged information~\cite{ma2025inference}.

\textbf{Oracle Verifier} assumes access to ground-truth reference signals and directly evaluates each candidate using a full-reference metric. In the context of audio SR, we adopt the \textbf{Log-Spectrogram Distance (LSD) Verifier} as the oracle verifier. LSD computes the L2 distance between the log-magnitude Short-Time Fourier Transform (STFT) spectrograms of the generated and reference audio. By operating in the log-spectral domain, it emphasizes perceptually salient differences. LSD is widely used in speech and audio restoration tasks as a perceptual proxy for audio fidelity~\cite{liu2024audiosr, liu2022neural, wang2021towards}.

\textbf{Supervised Verifiers} are employed under practical conditions where ground-truth reference signals are not accessible during inference. These verifiers consist of pretrained models that evaluate perceptual and semantic quality based on auxiliary conditioning inputs, rather than directly comparing against a reference signal.

For speech, we adopt the following verifiers:

\begin{itemize}[leftmargin=*]
    \item \textbf{Speaker Similarity (SpkSim) Verifier}: conditioned on a reference utterance from the target speaker, this verifier measures timbral consistency between the generated speech and the target speaker identity. It uses embeddings extracted from a pre-trained WavLM model~\cite{chen2022wavlm}.
    
    \item \textbf{Word Error Rate (WER) Verifier}: conditioned on target transcripts, it estimates transcription accuracy using a pre-trained automatic speech recognition (ASR) model~\cite{gao2023funasr, radford2023robust} and computes the edit distance between predicted and reference text.
    
    \item \textbf{Aesthetics (AES) Verifier}: operating without any external conditioning, the AudioBox-Aesthetics model~\cite{tjandra2025meta} offers no-reference audio quality assessment across four dimensions: Content Enjoyment (CE), Content Usefulness (CU), Production Complexity (PC), and Production Quality (PQ), providing a broad and interpretable evaluation for diverse audio types.
\end{itemize}

For non-speech audio such as music and sound effects, we utilize:

\begin{itemize}[leftmargin=*]
    \item \textbf{CLAP Verifier}: conditioned on textual descriptions, this verifier uses the Contrastive Language-Audio Pretraining (CLAP) model~\cite{elizalde2023clap} to assess semantic alignment between the generated audio and its associated caption.
    
    \item \textbf{Aesthetics (AES) Verifier}: as described above, applied to non-speech audio for assessing overall perceptual quality in a modality-agnostic fashion.
\end{itemize}

To reflect real-world scenarios where ground-truth references are unavailable, we primarily use supervised verifiers to guide the search, while reserving the oracle verifier for evaluation purposes. To improve robustness and mitigate the issue of \emph{verifier overfitting}—commonly referred to as \emph{verifier hacking}, where the generation process may overly adapt to a specific verifier’s scoring criteria, we further employ the \textbf{Ensemble Verifier}. This approach combines the feedback of all relevant supervised verifiers for a given audio category. Due to differences in scoring scales across verifiers, we adopt a rank-based aggregation strategy: for each sample, we compute its relative rank under each component verifier, then average these ranks to obtain a unified ensemble score. This strategy is also applied internally within the Aesthetics Verifier, where we aggregate the rank scores across CE, CU, PC, and PQ to produce a holistic quality estimate.

\subsection{Search Algorithms} 
Following the framework established in \cite{ma2025inference}, we adopt two representative search algorithms: \emph{Random Search} and \emph{Zero-Order Search}.

\textbf{Random Search} (as shown in Algorithm \ref{algo:random}) is the most straightforward strategy, implemented by sampling a set of $N$ initial Gaussian noises from an isotropic distribution. Each noise sample is then passed through the DDIM sampler to generate HR outputs. The top-1 result is selected according to the verifier score. 
While simple, this method is prone to \emph{verifier hacking}, as it explores the entire latent space without constraint, often exploiting verifier-specific biases~\cite{clark2023directly, ma2025inference, pan2022effects}.

\textbf{Zero-Order Search} (as shown in Algorithm \ref{algo:zero_order}) incorporates iterative refinement around the selected pivot noise. This process starts with a randomly sampled noise $\mathbf{n}$, and then explores a neighborhood around it. Formally, a local neighborhood is defined as 
\begin{equation}
S^{\lambda}_{\boldsymbol{n}, i} = \{ \mathbf{y} : d(\mathbf{y}, \mathbf{n}) = \lambda \}_{i=1}^K,    
\end{equation}
where $d(\cdot, \cdot)$ denotes a distance metric and $\lambda$ defines the distance of the search. From this neighborhood, $K$ candidates are generated and evaluated. The top-1 candidate is selected and then used as the new pivot noise. This process is repeated iteratively, gradually refining the search within a local region of the latent space.

\begin{figure}[t]
\centering
\vspace{-5mm}
\begin{minipage}{\linewidth}
\begin{algorithm}[H]
\caption{Random Search}
\label{algo:random}
\small
\begin{algorithmic}[1]
\State \textbf{Input:} Pretrained DM $\mathcal{D}_\theta$,Verifier $\mathcal{V}$,Number of Candidates $N$
\For{$i = 1$ to $N$}
    \State Sample noise $\boldsymbol{\epsilon}_i \sim \mathcal{N}(\boldsymbol{0}, \boldsymbol{I})$
    \State Generate sample $\boldsymbol{x}^{(i)} = \mathcal{D}_\theta(\boldsymbol{\epsilon}_i)$
    \State Evaluate score $\boldsymbol{s}^{(i)} = \mathcal{V}(\boldsymbol{x}^{(i)})$
\EndFor
\State Select top-1 output $\boldsymbol{x}^* = \arg\max_{i} \boldsymbol{s}^{(i)}$
\State \Return $\boldsymbol{x}^*$
\end{algorithmic}
\end{algorithm}
\end{minipage}
\hfill
\begin{minipage}{\linewidth}
\begin{algorithm}[H]
\caption{Zero-Order Search}
\label{algo:zero_order}
\small
\begin{algorithmic}[1]
\State \textbf{Input:} Initial noise $\boldsymbol{n}_0$, Search Distance $\lambda$, Neighbors $K$, \\ 
\qquad \quad Pretrained DM $\mathcal{D}_\theta$, 
Verifier $\mathcal{V}$, Number of Candidates $N$
\For{$i = 1$ to $N / K$}
    \For{$k = 1$ to $K$}
        \State Sample noise $\boldsymbol{\epsilon}^{(i)}$ at distance $\lambda$
        \State Generate sample $\boldsymbol{x}^{(i)} = \mathcal{D}_\theta(\boldsymbol{\epsilon}^{(i)})$
        \State Evaluate score $\boldsymbol{s}^{(i)} = \mathcal{V}(\boldsymbol{x}^{(i)})$
    \EndFor
    \State Update $\boldsymbol{n}_0 = \boldsymbol{\epsilon}^{(i^*)}, i^* = \arg\max_i \boldsymbol{s}^{(i)}$
\EndFor
\State \Return $\mathcal{D}_\theta(\boldsymbol{n}_0)$
\end{algorithmic}
\end{algorithm}
\end{minipage}
\end{figure}

\subsection{Search Space Range Estimation}

Existing research has yet to quantitatively characterize the \emph{range} of the search space induced by different search verifier-algorithm combinations, primarily due to limitations in task granularity and evaluation precision. The audio SR task presents a particularly suitable setting to bridge this gap, as the generated HR outputs are structurally aligned with the LR inputs, yet exhibit substantial perceptual variability.

For a given algorithm-verifier pair, let \( \mathcal{S}_N = \{ x^{(i)} \}_{i=1}^N \) denote a set of \( N \) generated HR outputs. We define the \textit{variance} of the search space as the average LSD  between the STFT of each candidate and the mean spectrogram of the set:
\begin{equation}
\left\{
\begin{aligned}
    \mu_N &= \frac{1}{N} \sum_{i=1}^{N} \text{STFT}(x^{(i)}) \\
    \text{Var}(\mathcal{S}_N) &= \frac{1}{N} \sum_{i=1}^{N} \text{LSD} \left[\text{STFT}(x^{(i)}), \mu_N \right]
\end{aligned}
\right.
\label{eq:search_space_range}
\end{equation}
This variance serves as a proxy for the search space’s diversity, reflecting the spread of plausible HR estimations under the given inference-time configuration.

\subsection{Uncertainty Estimation}

Recent advances in image SR have highlighted the value of uncertainty modeling. For example,~\cite{ning2021uncertainty} integrates spatial uncertainty into the training loss to enforce stronger supervision in ambiguous regions. Similarly,~\cite{zhang2025uncertainty} leverages deterministic models to estimate uncertainty via residuals between downsampled and upsampled images, guiding region-specific noise control for better reconstruction.
However, such residual-based strategies are less applicable to audio SR due to the absence of a well-defined HR spectrogram after downsampling and upsampling operations. As a result, residuals fail to capture the true ambiguity of HR estimation.

To address this, we propose to estimate uncertainty directly from the stochastic nature of diffusion sampling. We compute an \emph{uncertainty map} ~\cite{zhang2025uncertainty} by measuring the variance across time-frequency bins in the STFT domain over multiple generations from the same LR input. This approach reveals fine-grained regions of variability, providing insight into the ill-posedness of the task and the sensitivity of different spectral regions to sampling noise. Specifically, we estimate the variance at each time-frequency bin across all STFTs in \( \mathcal{S}_N \), and normalize the resulting values linearly:
\begin{equation}
\left\{
\begin{aligned}
    \mathcal{U}(t, f) &= \frac{\text{Var} \left[ \text{STFT}(x^{(i)})_{t, f} \right]_{i=1}^N - \min}{\max - \min + \epsilon}, \\
    \min &= \min \text{Var} \left[ \text{STFT}(x^{(i)})_{t, f} \right]_{i=1}^N, \\
    \max &= \max \text{Var} \left[ \text{STFT}(x^{(i)})_{t, f} \right]_{i=1}^N,
\end{aligned}
\right.
\end{equation}
where \( \mathcal{U}(t, f) \) denotes the normalized uncertainty score at time frame \( t \) and frequency bin \( f \), and \( \epsilon \) is a small constant for numerical stability. This formulation highlights regions with high generative variance, revealing time-frequency structures that are inherently ambiguous or sensitive to the generative stochasticity of diffusion models. 

\section{Experiments}
\label{sec:exp}

\begin{table*}[t]
\caption{
Comprehensive evaluation of inference-time scaling across verifier-algorithm combinations for speech, music, and sound effects at cutoff frequencies of 4\,kHz and 8\,kHz. \textbf{Bold} indicates top-1 performance, \underline{underline} denotes top-2, and \textit{italics} highlight cases where the performance of the generated output is worse than the LR input.
AES, SpkSim, WER, and LSD stand for Aesthetics Score, Speaker Similarity, Word Error Rate, and Log-Spectrogram Distance, respectively. The Ensemble Verifier aggregates AES, SpkSim, WER scores for speech, and AES, CLAP scores for music and sound effects. Random and Zero-Order denote \emph{Random Search} and \emph{Zero-Order Search}, respectively. 
}

\centering
\renewcommand{\arraystretch}{1.2}

% \subcaption*{(a) Speech}
\resizebox{0.7\textwidth}{!}{
\begin{tabular}{cl|cccc|cccc}
\toprule
\multicolumn{10}{c}{\textbf{Speech}} \\
\midrule
\multirow{2}{*}{} & \multirow{2}{*}{} 
& \multicolumn{4}{c|}{\textbf{4\,kHz}} 
& \multicolumn{4}{c}{\textbf{8\,kHz}} \\
\cmidrule(lr){3-6} \cmidrule(lr){7-10}
& & AES($\uparrow$) & SpkSim($\uparrow$) & WER($\downarrow$) & LSD($\downarrow$) & AES($\uparrow$) & SpkSim($\uparrow$) & WER($\downarrow$) & LSD($\downarrow$) \\

\midrule
\multicolumn{2}{c|}{LR Input} 
& 4.74 & 0.510 & 0.125 & 3.15 & 5.16 & 0.596 & 0.116 & 2.84 \\
\midrule
\multicolumn{2}{c|}{Vanilla AudioSR} 
& 4.74 & \textit{0.370} & \textit{0.263} & 1.73 & 5.18 & \textit{0.535} & \textit{0.136} & 1.62 \\
\midrule
\multirow{2}{*}{AES Verifier} 
  & + Random  & \textbf{5.23} & \textit{0.493} & \textit{0.144} & \underline{1.67} & \textbf{5.32} & \textit{0.572} & \textit{0.123} & \textbf{1.53} \\
  & + Zero-Order & 5.03 & \textit{0.458} & \textit{0.156} & \textbf{1.66} & 5.23 & \textit{0.570} & 0.116 & \textbf{1.53} \\
\midrule
\multirow{2}{*}{SpkSim Verifier} 
  & + Random  & 5.11 & \textbf{0.573} & \textit{0.161} & \underline{1.67} & 5.25 & \textbf{0.617} & 0.116 & 1.58 \\
  & + Zero-Order & 4.99 & \textit{0.471} & \textit{0.165} & \textbf{1.66} & 5.22 & \textit{0.582} & \textit{0.121} & \textbf{1.53}\\
\midrule
\multirow{2}{*}{WER Verifier} 
  & + Random  & 5.05 & \textit{0.463} & \textbf{0.099} & 1.70 & 5.25 & \textit{0.575} & \textbf{0.105} & 1.58 \\
  & + Zero-Order & 4.98 & \textit{0.458} & \textit{0.154} & \textbf{1.66} & 5.21 & \textit{0.575} & 0.114 & \textbf{1.53} \\
\midrule
\multirow{2}{*}{Ensemble Verifier} 

  & + Random  &  \underline{5.20}& \underline{0.540}
  & \underline{0.106} 
  & \underline{1.67} 
  & \underline{5.31} 
  & \underline{0.601}
  & \underline{0.106} 
  & \underline{1.55} \\
  & + Zero-Order & 5.02 & \textit{0.469} & \textit{0.152} & \textbf{1.66} & 5.23 & \textit{0.578} & 0.113 & \textbf{1.53} \\
\bottomrule
\end{tabular}  
}
\vspace{2mm}

\resizebox{0.92\textwidth}{!}{
\begin{tabular}{cl|ccc|ccc|ccc|ccc}
\toprule
\multirow{3}{*}{} & \multirow{3}{*}{} 
& \multicolumn{6}{c|}{\textbf{Music}} 
& \multicolumn{6}{c}{\textbf{Sound Effect}} \\
\cmidrule(lr){3-8} \cmidrule(lr){9-14}
& & \multicolumn{3}{c|}{\textbf{4\,kHz}} & \multicolumn{3}{c|}{\textbf{8\,kHz}} 
  & \multicolumn{3}{c|}{\textbf{4\,kHz}} & \multicolumn{3}{c}{\textbf{8\,kHz}} \\
\cmidrule(lr){3-5} \cmidrule(lr){6-8} \cmidrule(lr){9-11} \cmidrule(lr){12-14}
& & AES($\uparrow$) & CLAP($\uparrow$) & LSD($\downarrow$) 
  & AES($\uparrow$) & CLAP($\uparrow$) & LSD($\downarrow$) 
  & AES($\uparrow$) & CLAP($\uparrow$) & LSD($\downarrow$) 
  & AES($\uparrow$) & CLAP($\uparrow$) & LSD($\downarrow$) \\
\midrule
\multicolumn{2}{c|}{LR Input} & 6.06 & 0.340 & 3.95 & 6.47 & 0.352 & 3.09 & 4.16 & 0.458 & 3.97 & 4.42 & 0.481 & 3.33 \\
\midrule
\multicolumn{2}{c|}{Vanilla AudioSR} & 6.57 & \textit{0.303} & 2.20 & 6.63 & \textit{0.323} & 2.05 & 4.22 & \textit{0.343} & 2.86 & 4.43 & \textit{0.421} & 2.49 \\
\midrule
\multirow{2}{*}{AES Verifier} 
  & + Random     & \textbf{7.05} & 0.346 & 2.14 & \textbf{6.94} & 0.358 & \underline{2.03} & \textbf{4.72} & \textit{0.392} & 2.67 & \textbf{4.72} & \textit{0.455} & 2.49\\
  & + Zero-Order & 6.81 & 0.344 & 2.15 & 6.79 & 0.352 & \textbf{2.02} & 4.45 & \textit{0.398} & 2.76 & 4.54 & \textit{0.457} & \underline{2.44} \\
\midrule
\multirow{2}{*}{CLAP Verifier} 
  & + Random     & 6.80 & \textbf{0.414} & \underline{2.12} & 6.75 & \textbf{0.424} & 2.04 & 4.39 & \textbf{0.495} & 2.77 & 4.51 & \textbf{0.522} & 2.53 \\
  & + Zero-Order & 6.77 & 0.357 & 2.16 & 6.77 & 0.373 & 2.05 & 4.39 & \textit{0.409} & 2.74 & 4.50 & \textit{0.471} & \textbf{2.43} \\
\midrule
\multirow{2}{*}{Ensemble Verifier} 
  & + Random     & \underline{6.97} & \underline{0.394} & \textbf{2.09} & \underline{6.89} & \underline{0.404} & \underline{2.03} & \underline{4.62} & 0.464 & \textbf{2.65} & \underline{4.65} & \underline{0.502} & 2.48 \\
  & + Zero-Order & 6.79 & 0.352 & 2.18 & 6.78 & 0.368 & \underline{2.03} & 4.41 & \textit{0.415} & 2.72 & 4.54 & \textit{0.469} & \textbf{2.43} \\
\bottomrule
\end{tabular}
}
\label{tab:overall}
\end{table*}

\begin{table*}[th]
\centering
\caption{Comparison of search range across algorithms based on LSD variance for different audio types at 4\,kHz, 8\,kHz, and their average. Random and Zero-Order denote \emph{Random Search} and \emph{Zero-Order Search}, respectively.}
\renewcommand{\arraystretch}{1.2}
\resizebox{0.6\textwidth}{!}{
\begin{tabular}{l|ccc|ccc|c}
\toprule
\multirow{2}{*}{\textbf{Algorithm}} & \multicolumn{3}{c|}{\textbf{4\,kHz}} & \multicolumn{3}{c|}{\textbf{8\,kHz}} & \textbf{Average} \\
 & Speech & Music & Sound Effect & Speech & Music & Sound Effect & \textbf{LSD Variance} \\
\midrule
Random  & \textbf{0.673} & \textbf{0.921} & \textbf{1.25} & \textbf{0.577} & \textbf{0.868} & \textbf{1.10} & \textbf{0.898} \\
Zero-Order & 0.608 & 0.790 & 0.947 & 0.529 & 0.750 & 0.940 & 0.761 \\
% Paths-2 & 0.597 & 0.792 & 1.02 & 0.524 & 0.619 & 1.02 & 0.762 \\
\bottomrule
\end{tabular}
}
\label{tab:search_range_variance}
\end{table*}

\subsection{Evaluation Datasets}

We construct three curated evaluation benchmarks, each comprising 200 samples, to assess model performance across different audio types: VCTK \cite{liu2022neural} for speech, MusicCaps \cite{agostinelli2023musiclm} for music, and ESC-50 \cite{piczak2015esc} for sound effects. These datasets are tailored to reflect the specific characteristics and evaluation criteria of each audio domain. The detailed settings for each audio type are as follows.

For \emph{Speech}, to evaluate intelligibility and speaker consistency, we use the ground-truth transcripts provided by VCTK to compute the Word Error Rate (WER). Additionally, we randomly sample other utterances from the same speaker to serve as audio references for the Speaker Similarity (SpkSim) Verifier.
For \emph{Music}, to ensure evaluation quality, we first pre-filter low-quality samples using MusicCaps captions by discarding entries containing keywords such as \textit{``mediocre''}, \textit{``low quality''}, or \textit{``low fidelity''}. From the remaining subset, we select samples that span diverse genres and exhibit rich high-frequency content, thereby better testing the SR capability.
For \emph{Sound Effect}, since ESC-50 lacks human-written descriptions, we generate audio captions using Qwen2-Audio \cite{chu2024qwen2}, conditioned on the original ESC-50 category keywords. These synthesized captions are then paired with their corresponding audio clips and used in CLAP Verifier, which evaluates semantic alignment via contrastive audio-text embeddings.

\subsection{Search Settings}
\label{sec:search_settings}

For AudioSR, we retain its default settings using 50 DDIM sampling steps and classifier-free guidance scale of 3.5 during inference-time sampling.
We fix the size of the inference-time search space to \( N = 120 \) for each algorithm-verifier pair. For each configuration, we select the top-1 HR output according to the verifier scores.
We denote \emph{Random Search} and \emph{Zero-Order Search} as \textbf{Random} and \textbf{Zero-Order}, respectively. The \emph{Zero-Order Search} algorithm is configured with \( K = 2 \) neighborhood candidates and a search distance parameter of \( \lambda = 0.99 \).

%\begin{figure}[th]
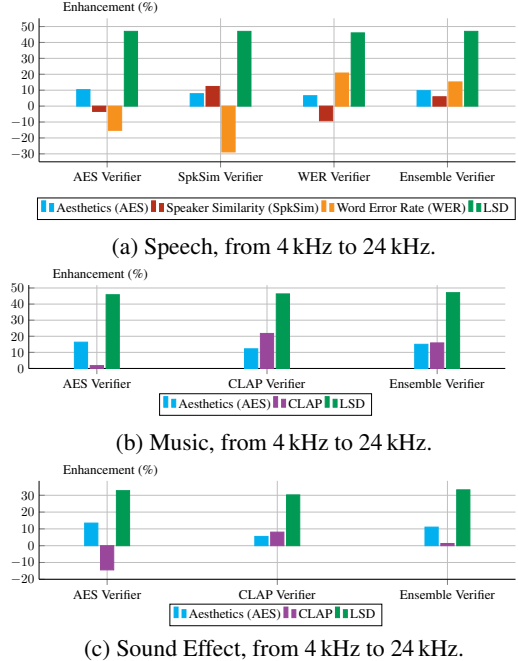
\begin{wrapfigure}{r}{0.51\textwidth}
  \vspace{-1.5cm}
  \begin{subfigure}{\linewidth}
    \begin{tikzpicture}[scale=0.5]
    \begin{axis}[
    ybar,ymin=-35,
    ytick={-30, -20, -10, 0, 10, 20, 30, 40, 50},
    width=2.0\linewidth, % Adjust the width as needed
    enlarge x limits=0.2,
    height=5.4cm,
    ylabel=Enhancement (\%),
    % x tick label style={rotate=0, anchor=east, align=center},
    legend style={at={(1.0, -0.25), anchor=north}, legend columns=-1},
    symbolic x coords={AES Verifier, SpkSim Verifier, WER Verifier, Ensemble Verifier},
    y label style={at={(axis description cs:0.15,1.0)}, anchor=south, rotate=270}, % Adjust the position and font size
    xtick=data,
    ymajorgrids=true,
    xmajorgrids=true,
    axis lines*=left,
    ]
    \addplot +[draw=ProcessBlue,fill=ProcessBlue] coordinates {(AES Verifier, 10.34) (SpkSim Verifier, 7.81) (WER Verifier, 6.54) (Ensemble Verifier, 9.70)};
    \addlegendentry{Aesthetics (AES)}
    
    \addplot +[draw=BrickRed,fill=BrickRed] coordinates {(AES Verifier, -3.33) (SpkSim Verifier, 12.35) (WER Verifier, -9.21) (Ensemble Verifier, 5.88)};
    \addlegendentry{Speaker Similarity (SpkSim)}
    
    \addplot +[draw=BurntOrange, fill=BurntOrange] coordinates {(AES Verifier, -15.20) (SpkSim Verifier, -28.80) (WER Verifier, 20.80) (Ensemble Verifier, 15.20)};
    \addlegendentry{Word Error Rate (WER)}
    
    \addplot +[draw=ForestGreen, fill=ForestGreen] coordinates {(AES Verifier, 46.98) (SpkSim Verifier, 46.98) (WER Verifier, 46.03) (Ensemble Verifier, 46.98)};
    \addlegendentry{LSD}
    \end{axis}
    \end{tikzpicture}
    
    \caption{Speech, from 4\,kHz to 24\,kHz. }
    \label{fig:short-a}      
  \end{subfigure}
  
  \begin{subfigure}{\linewidth}
    \begin{tikzpicture}[scale=0.5]
    \begin{axis}[
    ybar,ymin=0,
    ytick={0, 10, 20, 30, 40, 50},
    width=2.0\linewidth, % Adjust the width as needed
    enlarge x limits=0.2,
    height=3.8cm,
    ylabel=Enhancement (\%),
    % x tick label style={rotate=0, anchor=east, align=center},
    legend style={at={(0.72, -0.3), anchor=north}, legend columns=-1},
    symbolic x coords={AES Verifier, CLAP Verifier, Ensemble Verifier},
    y label style={at={(axis description cs:0.15,1.0)}, anchor=south, rotate=270}, % Adjust the position and font size
    xtick=data,
    ymajorgrids=true,
    xmajorgrids=true,
    axis lines*=left,
    ]
    \addplot +[draw=ProcessBlue,fill=ProcessBlue] coordinates {(AES Verifier, 16.34) (CLAP Verifier, 12.21) (Ensemble Verifier, 15.02)};
    \addlegendentry{Aesthetics (AES)}
    
    \addplot +[draw=Purple, fill=Purple] coordinates {(AES Verifier, 1.76) (CLAP Verifier, 21.76)  (Ensemble Verifier, 15.88)};
    \addlegendentry{CLAP}
    
    \addplot +[draw=ForestGreen, fill=ForestGreen] coordinates {(AES Verifier, 45.82) (CLAP Verifier, 46.33) (Ensemble Verifier, 47.09)};
    \addlegendentry{LSD}
    \end{axis}
    \end{tikzpicture}
    
    \caption{Music, from 4\,kHz to 24\,kHz. }
    \label{fig:short-b}      
  \end{subfigure}

  \begin{subfigure}{\linewidth}
    \begin{tikzpicture}[scale=0.5]
    \begin{axis}[
    ybar,ymin=-20,
    ytick={-20, -10, 0, 10, 20, 30, 40},
    width=2.0\linewidth, % Adjust the width as needed
    enlarge x limits=0.2,
    height=4.2cm,
    ylabel=Enhancement (\%),
    % x tick label style={rotate=0, anchor=east, align=center},
    legend style={at={(0.72, -0.25), anchor=north}, legend columns=-1},
    symbolic x coords={AES Verifier, CLAP Verifier, Ensemble Verifier},
    y label style={at={(axis description cs:0.15,1.0)}, anchor=south, rotate=270}, % Adjust the position and font size
    xtick=data,
    ymajorgrids=true,
    xmajorgrids=true,
    axis lines*=left,
    ]
    \addplot +[draw=ProcessBlue,fill=ProcessBlue] coordinates {(AES Verifier, 13.46) (CLAP Verifier, 5.53) (Ensemble Verifier, 11.06)};
    \addlegendentry{Aesthetics (AES)}
    
    \addplot +[draw=Purple, fill=Purple] coordinates {(AES Verifier, -14.41) (CLAP Verifier, 8.08)  (Ensemble Verifier, 1.31)};
    \addlegendentry{CLAP}
    
    \addplot +[draw=ForestGreen, fill=ForestGreen] coordinates {(AES Verifier, 32.75) (CLAP Verifier, 30.23) (Ensemble Verifier, 33.25)};
    \addlegendentry{LSD}
    \end{axis}
    \end{tikzpicture}
    \vspace{-0.6mm}
    \caption{Sound Effect, from 4\,kHz to 24\,kHz. }
    \label{fig:short-c}      
  \end{subfigure}
  \caption{
  Performance improvements over the LR input across different audio types using inference-time \emph{Random Search} with various verifiers from 4\,kHz to 24\,kHz. The \emph{enhancement} denotes the relative improvement over LR under each evaluation metric. For Aesthetics (AES), Speaker Similarity (SpkSim) and CLAP Score, enhancements reflect relative increases. For Word Error Rate (WER) and Log Spectrogram Distance (LSD), improvements are computed as relative reductions. 
  }
  \label{fig:perf_gain}
  \vspace{-1cm}
%\end{figure}
\end{wrapfigure}

\subsection{Experiment Result Analysis}

\paragraph{Effectiveness of Scaling Inference-Time Compute.}

As shown in Table~\ref{tab:overall} and Figure~\ref{fig:perf_gain}, we present a comprehensive evaluation of performance improvements across verifier-algorithm combinations for speech, music, and sound effects at cutoff frequencies of 4\,kHz and 8\,kHz. While vanilla audio SR improves perceptual fidelity in general, it often comes at the cost of degrading critical attributes when compared to the LR input. For speech, this degradation is particularly evident in metrics such as Speaker Similarity and WER, which reflect speaker consistency and intelligibility. Similarly, for music and sound effects, we observe a consistent drop in CLAP Score, which measures the alignment between audio and semantic content. These findings underscore the limitations of vanilla SR generation and motivate the use of verifier-guided inference-time scaling strategies to restore such domain-specific qualities.
Crucially, we find that when the verifier used for search directly aligns with the evaluation metric—such as using the Speaker Similarity Verifier to optimize Speaker Similarity, or the CLAP Verifier to improve CLAP Score—\emph{Random Search} consistently outperforms \emph{Zero-Order Search} in recovering the lost performance. This is largely due to its ability to explore a broader candidate space, enabling more effective correction of deficiencies introduced by the vanilla SR process.

\paragraph{Search Space Range Estimation.}

Empirical evidence suggests that \emph{Random Search} has a higher likelihood of locating global optima, whereas \emph{Zero-Order Search} exhibits stronger locality due to its iterative refinements around selected initial noise samples \cite{ma2025inference}. To quantitatively assess this behavior, we estimate the search space range using Equation~\ref{eq:search_space_range}. As shown in Table~\ref{tab:search_range_variance}, the search space range tends to be larger for lower cutoff frequencies compared to their higher-frequency counterparts. Moreover, we observe a progressive increase in search space range from speech to music and then to sound effects, suggesting rising levels of uncertainty across these audio types. Consistently, \emph{Random Search} exhibits a significantly wider search range than \emph{Zero-Order Search}, confirming its advantage in exploring diverse generative candidates.

\begin{figure*}[tb]
  \centering
  \includegraphics[width=\linewidth]{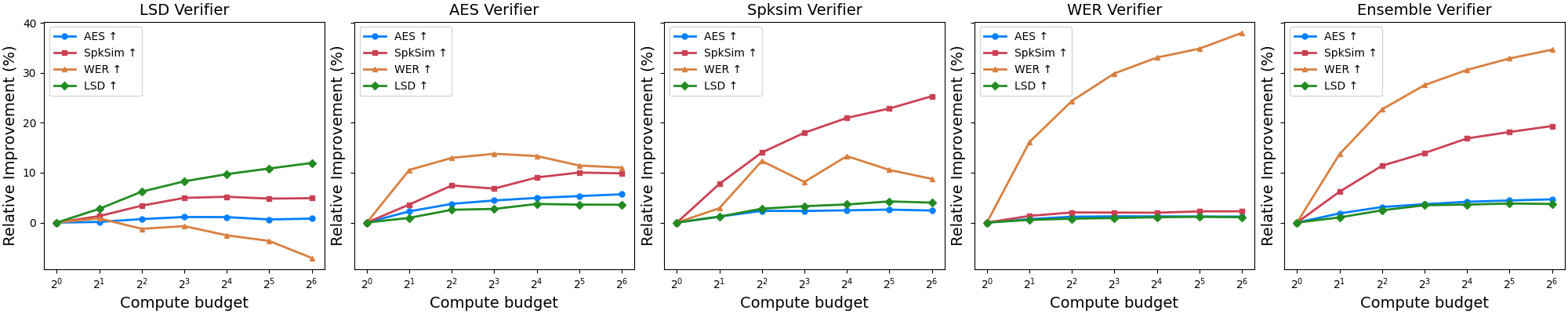}
  \caption{
    Relative performance improvements over the default generation (vanilla AudioSR) for speech from 4\,kHz to 24\,kHz in \emph{Random Search}, demonstrating the effect of inference-time scaling across different verifier types. 
    LSD, AES, SpkSim, and WER refer to Log Spectrogram Distance, Aesthetics Score, Speaker Similarity, and Word Error Rate, respectively. 
    The Ensemble Verifier aggregates AES, SpkSim, and WER by averaging their rank scores. 
  }
  %     Among these, LSD serves as the \emph{Oracle Verifier}, while AES, SpkSim, and WER are treated as \emph{Supervised Verifiers}.
  \label{fig:scale_up}
  \vspace{-3mm}
\end{figure*}

\paragraph{Uncertainty Estimation. }

\begin{wrapfigure}{r}{0.55\textwidth}
%\begin{figure}[tb]
  \vspace{-1.5mm}
  \centering
  %\begin{subfigure}{0.9\linewidth}
    \includegraphics[width=\linewidth]{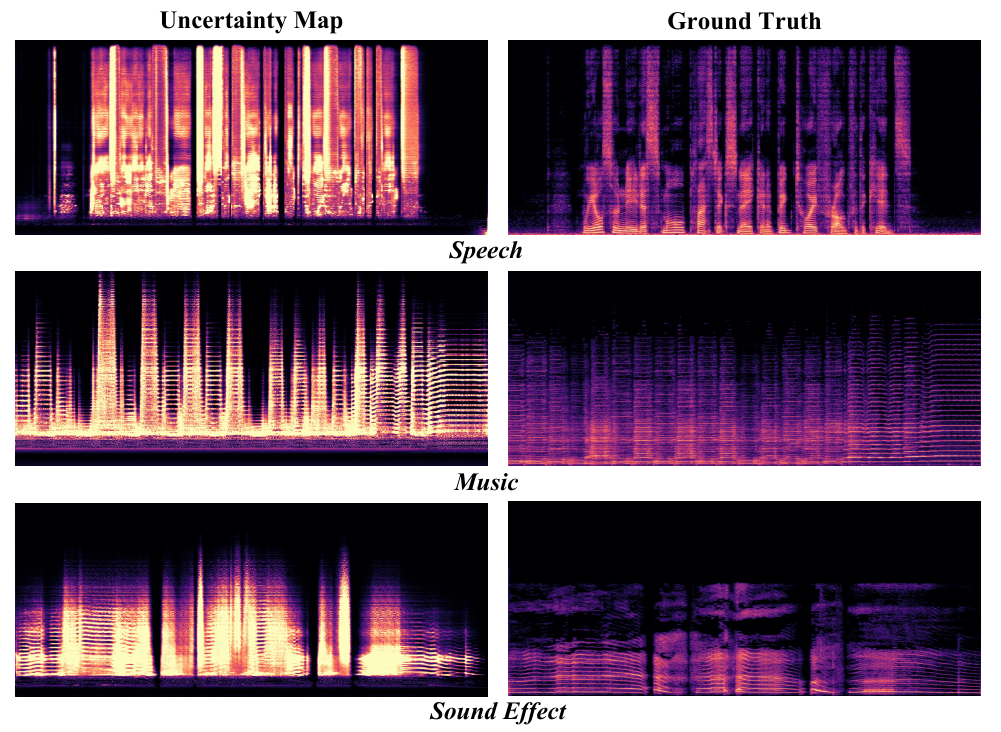}
  %\hfill
  \caption{Visualization of uncertainty maps over STFT spectrograms across diverse audio types. Brighter regions indicate higher generative uncertainty across diffusion samples. Overlapping high-uncertainty areas reflect divergent spectral realizations, underscoring the necessity of inference-time scaling to identify perceptually optimal outputs.}
  \vspace{-1cm}
\label{fig:uncertainty}
%\end{figure}
\end{wrapfigure}

One of the central challenges in audio SR with diffusion models stems from the inherent randomness of the denoising process, which leads to significant output variability across different sampling runs. However, this stochastic behavior is often under-characterized. To address this, we employ uncertainty map visualization as an interpretability tool to highlight the localized variance patterns embedded within individual samples, as shown in Figure \ref{fig:uncertainty}.
For implementation, we compute the variance across the time-frequency bins of STFT spectrograms derived from \emph{Random Search} generation candidates, since this algorithm has a wider search range. To enhance visual interpretability, we apply percentile-based contrast normalization by clipping values above the 90th percentile, thus emphasizing the structure of salient uncertainty regions prior to rendering.

\paragraph{Verifier Hacking. }

To examine the implications of scaling up inference-time compute, we focus our analysis on \emph{Random Search}, which offers a wider search range. This setting allows us to better isolate the phenomenon of \emph{verifier hacking}, where the optimization overfits to a specific verifier without achieving meaningful or comprehensive improvements. As shown in Figure~\ref{fig:scale_up}, we present the relative performance improvements while scaling up inference-time compute budget in \emph{Random Search} over the default generation in speech from 4\,kHz to 24\,kHz. Among the verifiers, LSD Verifier serves as \emph{Oracle Verifier}, with an access to ground-truth references while searching. Aesthetics (AES), Speaker Similarity (SpkSim), Word Error Rate (WER), CLAP and Ensemble Verifiers are treated as \emph{Supervised Verifiers}, which we employ at practical settings without referring to ground-truth signals while searching. 

Among the evaluation metrics for speech, WER behaves uniquely. Notably, when guided by the \emph{Oracle Verifier}, LSD Verifier, we observe that higher LSD scores do not necessarily correspond to improvements in essential speech attributes—particularly WER. In fact, WER often worsens even as LSD improves, revealing a misalignment between general fidelity and intelligibility. This highlights the need for \emph{Supervised Verifiers} during inference-time search to preserve perceptual quality specific to the speech SR task.

Furthermore, while the WER Verifier is effective in improving WER alone, it provides limited gains in other metrics, emphasizing its narrow focus. As we scale up the inference-time compute budget, we observe that both the Aesthetics and Speaker Similarity Verifiers begin to exhibit diminishing or even negative returns on WER beyond a search space size of approximately \( 2^3 \sim 2^4 \), indicating a \emph{verifier hacking} phenomenon, where the search process overfits to the verifier at the cost of overall quality.
In contrast, the \emph{Ensemble Verifier} mitigates such overfitting by averaging the ranks from multiple verifiers (AES, SpkSim, and WER), leading to a more balanced trade-off across all metrics and achieving meaningful improvements in WER without compromising other perceptual qualities.

\paragraph{Verifier-Algorithm-Task Alignment.}

When upsampling from 4\,kHz to 24\,kHz, employing the \emph{Ensemble Verifier} with \emph{Random Search} consistently yields the best trade-off across all audio types. As the SR setting shifts from a 4\,kHz to an 8\,kHz input (both upsampled to 24\,kHz), different audio types begin to exhibit distinct preferences for specific verifier-algorithm configurations. For speech, the combination of \emph{Ensemble Verifier} and \emph{Random Search} continues to outperform both the LR input and vanilla SR baseline, stemming from its ability to jointly enhance holistic speech-related attributes, demonstrating robust trade-off performance across evaluation metrics.

Interestingly, at the 8\,kHz upsampling setting, \emph{Zero-Order Search} achieves relatively better LSD scores than \emph{Random Search}. This is because the sample variance decreases when starting from higher-resolution inputs, reducing the benefit of a wider search space. In such scenarios, when metrics like Speaker Similarity and CLAP Score fluctuate within narrow margins, \emph{Zero-Order Search} proves advantageous by mitigating the risk of \emph{verifier hacking} and generating outputs that better align with ground-truth signals. Moreover, at 8\,kHz, we observe that music shows a stronger preference for the Aesthetics Verifier, while sound effects align better with the CLAP Verifier—underscoring the importance of verifier-algorithm-task alignment in optimizing performance.

\section{Conclusion}

This work presents a comprehensive study of inference-time scaling for diffusion models in the context of audio SR.
We propose a unified framework that explores scalable search strategies guided by diverse verifier-algorithm combinations, achieving consistent improvements across audio types and cutoff frequencies. We further identify the phenomenon of verifier hacking and demonstrate that verifier ensembling effectively mitigates this issue by balancing competing perceptual objectives.
Beyond performance gains, we quantify search space ranges via LSD variance, revealing how search dynamics vary across domains and upsampling settings. We also introduce variance-based uncertainty maps to highlight time-frequency regions sensitive to generative noise, offering deeper insights into the stochasticity and ill-posedness of the audio SR task, while opening up promising directions for future research in uncertainty-aware diffusion modeling in audio generation.

%\newpage
\bibliography{main}
\bibliographystyle{plain}

\end{document}